\documentclass[12pt]{article}
\usepackage{amssymb,latexsym}
\usepackage[dvips]{graphicx}
\def\beq{\begin{equation}}
\def\eeq{\end{equation}}
\def\bea{\begin{eqnarray}}
\def\eea{\end{eqnarray}}
\def\beas{\begin{eqnarray*}}
\def\eeas{\end{eqnarray*}}
\def\nn{\nonumber}

\def\N{\mathbb{N}} 
 
\def\P{{\cal P}}
\def\ra{\rangle}

\def\l{\ldots}
\def\c{\cdots}

\def\e{\epsilon}
\def\q{{\quad}}

\renewcommand{\theequation}{\arabic{section}.\arabic{equation}}
\begin{document}
%
%
\begin{center} 
{\Large \bf 
Macroscopic properties of $A$-statistics}\\[5mm]
{\bf A.\ Jellal}\footnote{E-mail~: jellal@gursey.gov.tr.}\\[1mm]
Feza Gursey Institute,\\
P.O.~Box 6, 81220 Cengelkoy, Istanbul, Turkey;\\[2mm]
{\bf T.D.\ Palev}\footnote{E-mail~: tpalev@inrne.bas.bg.}\\[1mm]
Institute for Nuclear Research and Nuclear Energy,\\ 
Boul.\ Tsarigradsko Chaussee 72, 1784 Sofia, Bulgaria;\\[2mm] 
{\bf and J.\ Van der Jeugt}\footnote{E-mail~:
Joris.VanderJeugt@rug.ac.be.}\\[1mm]
Department of Applied Mathematics and Computer Science, 
University of Ghent,\\
Krijgslaan 281-S9, B-9000 Gent, Belgium. 
\end{center}


\vskip 10mm
\begin{abstract}
$A$-statistics is defined in the context of the Lie algebra $sl(n+1)$.
Some thermal properties of $A$-statistics are
investigated under the assumption that the particles interact
only via statistical interaction imposed by the Pauli principle of
$A$-statistics. Apart from the general case, three particular examples
are studied in more detail~: (a) the particles have one and the same
energy and chemical potential; (b) equidistant energy spectrum;
(c) two species of particles with one and the same energy and
chemical potential within each class. The grand partition functions and
the average number of particles are among the thermodynamical quantities
written down explicitly.
\end{abstract} 
\vskip 2mm

\section{Introduction}
\setcounter{equation}{0}

The first attempts to generalize canonical quantum
statistics go back to Gentile~\cite{Gentile}, who considered 
statistics, which are intermediate between Fermi-Dirac (FD) and
Bose-Einstein (BE) statistics. More precisely, Gentile introduced
statistics with the property that the maximal
occupation number of particles on any orbital is
larger than~1 (hence the statistics is not FD), but is finite 
(hence the statistics is not BE). 
Since that time various generalizations of quantum
statistics have been proposed both in quantum field 
theory~\cite{Wigner,Green,Greenberg}
and in condensed matter physics~\cite{Wilczek,Haldane,Wu}, 
some of them inspired by new developments in 
conformal field theories and related lattice models
(~\cite{Berkovich} and references therein) and in
quantum groups~\cite{Pusz,Macfarlane}.
For an overview of generalized quantum statistics formulated in
terms of deformed algebras or generalized Fock spaces,
we refer to~\cite{Bonatsos,Mishra}.

In 1950, Wigner~\cite{Wigner} has shown (on a simple example) 
that there might exist statistics, which are compatible with 
the principles of quantum theory without the necessity that 
the position and the momentum operators 
satisfy the canonical commutation relations. 
This more general statistics
discovered by Wigner turned out to be the para-Bose statistics 
of one pair of creation and annihilation operators (CAO's)~\cite{OK}. 
Three years later Green introduced both para-Bose (pB) and 
para-Fermi (pF) statistics in the more general frame of 
quantum field theory~\cite{Green}.

In the present paper we study the macroscopic properties 
of a certain type of statistics, called $A$-statistics. 
It was introduced in~\cite{Palev76,Palev77} and studied further 
from the microscopic point of view in~\cite{PV00}.
$A$-statistics resembles the pF statistics insofar as
the creation and the annihilation operators of both statistics 
generate simple Lie algebras~: 
any $n$ pairs of parafermions generate the
orthogonal Lie algebra $so(2n+1)\equiv B_n$~\cite{KT,RS}, 
whereas any $n$ pairs of $A$-CAO's generate the Lie algebra 
$sl(n+1)\equiv A_n$ (which justifies
the name $A$-statistics). 
$A$-statistics resembles also Bose
statistics~: similar to bosons, the $A$-creation (resp.\ 
annihilation) operators commute with each other. 
The Fock representations for pF, pB and $A$-statistics
are constructed in one and the same way~: they are
generated out of a vacuum by creation operators only.
The Fock representations in all three cases
are labelled by a positive integer $p=1,2,\l$, called
the order of statistics. 
Moreover the metric within any Fock space is defined 
with the usual Fock space technique.
It is essential to point out that contrary to the CAO's
of parastatistics, the $A$-creation operators $a_1^+$, $\ldots$,
$a_n^+$ (resp.\ the $A$-annihilation operators $a_1^-$, $\ldots$,
$a_n^-$) commute with each other. For this reason (apart from
the trivial case) they differ essentially from the CAO's
of the $g$-ons~\cite{Karabali} or from the CAO's associated
with solutions of the spectral Yang-Baxter equations~\cite{Liguori},
since in these works relations of the type 
$a_i^-a_j^-=R_{ij} a_j^- a_i^-$ are imposed.

In the case of para-Fermi statistics of order $p$ no more than 
$p$ particles can be accommodated on any orbital. 
The filling of the orbitals is
however completely independent of each other. 
Here comes one of the essential differences with $A$-statistics. 
The Pauli principle for $A$-statistics says that if the order of
statistics is $p$, then the system cannot accommodate more than $p$
particles. Thus, if $p=10$ and 10 particles are already accommodated
on the first orbital, then no more particles can be added to any
orbital. For this reason $A$-statistics gives perhaps the simplest 
example of an exclusions statistics~\cite{Haldane,Wu}~: 
the number of the available places on a certain orbital depends 
on how many particles (independently where) are already accommodated  
in the system (see~\cite{PV00} for more discussions of this issue). 
Here it is an appropriate place to say that the word {\em particle} 
is used in the context of this paper as a collective name for particles, 
quasiparticles, excitations, etc.

In Section~2 we recall shortly the definition and the main 
microscopic properties of $A$-statistics. Similarly as for para-Fermi 
statistics the creation and the annihilation operators $a_1^\pm,\l,a_n^\pm$
of $sl(n+1)$ are defined via triple commutation relations (see (\ref{2.1})). 
These triple relations define completely the Lie algebra $sl(n+1)$, a
property which was indicated for the first time by Jacobson~\cite{Jacobson}. 
For this reason we call the CAO's of $A$-statistics {\em Jacobson generators}.

In Section~3 we write down explicitly the $sl(n+1)$ grand partition 
function $Z(p,n)$ and the average number of particles in the system 
${\bar N}(p,n)$, see equations (\ref{3.9}) and (\ref{3.16}),
under the general assumption that the energy of each particle 
on orbital~$i$ is $\e_i$. In this context $n$ is the number of 
orbitals of the system and $p$ is the order of the statistics, 
a positive integer, which labels the inequivalent Fock space 
representations, see (\ref{2.4}). Because of the Pauli principle 
the orbitals cannot be considered as independent subsystems
(as in BE or FD statistics)~: 
the filling of any orbital depends on the states of the other orbitals. 
Therefore we derive the thermodynamical quantities directly for
the $n$-orbital system, assuming that it is in a thermal and diffusive
contact and in a thermal and diffusive equilibrium with a much bigger 
reservoir. As we shall see, the $k$-th complete symmetric functions, 
see (\ref{3.6}), turn out to be a particularly convenient tool
for the description of the thermal properties of the system.

In the remaining three sections we consider different specializations of
the general settings of Section~3. First (Section~4) we assume that all 
orbitals (i.e.\ single particle states) have one and the same energy and
chemical potential. We express the grand partition function 
and the ensemble average number of particles via hypergeometric 
functions (see, for instance, equations (\ref{3.32}) and (\ref{3.39b})). 
Two special cases are considered in some more detail. 
The first one corresponds to $n=1$. Here, the $p=1$ representation leads to
the Fermi-Dirac distribution function, whereas $p=\infty$ corresponds 
to the Bose-Einstein distribution. For all other values of $p$ 
the distribution function is intermediate between the FD and BE 
distributions. The second case, corresponding to $p=1$, see Figure~2, 
leads to the so-called hard-core fermions or hard-core bosons 
(locally they coincide). Such particles are natural ingredients 
in multi-band Hubbard or various Heisenberg spin models, 
where configurations which contain more than one particle on each 
lattice site are strictly prohibited (see~\cite{PV00} for more 
discussions on the topic).

In Section~5 a model with equidistant energy levels is considered.
The orbitals are labelled by the energy. The grand partition function 
is written in terms of the so-called $q$-generalized or basic 
hypergeometric functions, see (\ref{3.64}). The conclusion is that for big
energy gaps or at very low temperatures all particles ``condensate"
on the lowest energy orbital. 
The case with $p=1$ is considered in more details.

In Section~6 we consider two species of particles. Those of the 
first kind $A$ (resp.\ of kind B) have one and the same energy 
$\e_A$ (resp.\ $\e_B$)
and chemical potential $\mu_A$ (resp.\ $\mu_B$).
Apart from the grand partition function and the average number 
of particles ${\bar N}(p,n)$, also the thermal average 
${\bar N}(p,n)_A$  of the number of particles of kind~$A$ and 
of kind~$B$ are computed. On the example of $sl(5)$ with $p=4$ 
the general accommodation properties are demonstrated. 
For instance the region with $\e_A < \mu_A$ and
$\e_B > \mu_B$ is populated most probably with particles 
of the first kind, see Figure~4, whereas the region with $\e_A < \mu_A$ and
$\e_B < \mu_B$ is populated with approximately the same 
number of particles of both kinds, see Figure~3.

\vskip 2mm
Throughout the paper we use the following abbreviations and
notation (some of them standard)~:
\begin{itemize}
\item[] CAO's~: creation and annihilation operators;
\item[] GPF~: grand partition function;
\item[] $\N$~: all positive integers;
\item[] $[a,b]=ab-ba$.
\end{itemize}

\section{Microscopic properties of $A$-statistics}
\setcounter{equation}{0}

In this section we list shortly the basic definitions and some of the
microscopic properties of $A$-statistics. 
In particular, we shall define 
\begin{itemize}
\item the CAO's of $A$-statistics and their ``triple 
commutation'' relations,
\item the Fock spaces of $A$-statistics, and the corresponding Pauli
principle,
\item the Hamiltonian being studied in these Fock spaces.
\end{itemize}
For more details and a derivation of the results we refer 
to~\cite{Palev76,Palev77,PV00}.

The CAO's of $A$-statistics are equal to
the Jacobson creation and annihilation operators 
$a_1^\pm, a_2^\pm, \l, a_n^\pm$ of $sl(n+1)$, which are
defined as $2n$ operators satisfying the relations
\bea
&& [[a_i^+,a_j^-],a_k^+]
   =\delta_{kj}a_i^+ +\delta_{ij}a_k^+ ,\nn\\
&& [[a_i^+,a_j^-],a_k^-]=-\delta_{ki}a_j^- - \delta_{ij}a_k^-,
\label{2.1}\\
&& [a_i^+,a_j^+]=[a_i^-,a_j^-]=0. \nn
\eea
The $sl(n+1)$ generators expressed in terms of the Jacobson CAO's read~:
\beq
e_{i0}=a_i^+,\quad e_{0i}=a_i^-, \quad
e_{ii}-e_{00}=[a_i^+,a_i^-],\quad
e_{ij}=[a_i^+,a_j^-];\quad i\ne j=1,\l,n. 
\label{2.2}
\eeq
Above $\{e_{ij}|i,j=0,1,\ldots,n\}$ are the known Weyl generators
of $gl(n+1)$~:
\beq
[e_{ij},e_{kl}]=\delta_{jk}e_{il}-\delta_{il}e_{kj}.\label{2.3}
\eeq

As in the case of parastatistics~\cite{Green} the Fock spaces $W(p,n)$ of 
$A$-statistics are labelled by an order of statistics~$p$,
where $p$ runs over all positive integers~: $p\in \N$.
Each state space $W(p,n)$ is defined by the
requirement that it contains a vector $|0\ra$, a vacuum,
such that
\beq
a_i^-a_j^+|0\ra=\delta_{ij}p|0\ra,\qquad 
a_k^-|0\ra=0;\qquad p\in \N,
\quad i,j,k=1,\l,n. \label{2.4}
\eeq
The Fock spaces are finite-dimensional irreducible $sl(n+1)$-modules.
All vectors
\beq
(a_1^+)^{l_1}(a_2^+)^{l_2}\l(a_n^+)^{l_n}|0\ra \label{2.5}
\eeq
subject to the restriction
\beq
 l_1+l_2+\c+l_n\le p \label{2.6}
\eeq
constitute a basis in $W(p,n)$.

Here a remark is in order. The linear span of all vectors~(\ref{2.5})
for any $l_1,\l,l_n \in \{0,1,2,\ldots\}$, namely 
without the restriction~(\ref{2.6}), 
is an infinite-dimensional $sl(n+1)$-module $\tilde{W}(p,n)$. 
The latter is however not irreducible. $\tilde{W}(p,n)$ 
contains an (infinite-dimensional) invariant subspace $W_{inv}(p,n)$,
which is the linear envelope of all vectors~(\ref{2.5}) with
$l_1+l_2+\l+l_n > p$. Then $W(p,n)$ is a factor module of 
$\tilde{W}(p,n)$ with respect to $W_{inv}(p,n)$
(and the vectors~(\ref{2.5}) subject to the restriction~(\ref{2.6}) 
are representatives of the corresponding equivalent classes in 
$\tilde{W}(p,n)/W_{inv}(p,n)$).

Define a Hermitian form $(~,~)$ on $W(p,n)$ with the usual Fock space 
technique, namely postulating (in addition to $a_i^-|0\ra=0$) that 
\bea
&\hbox{(a)}& \langle 0|0\ra=1, \nn\\
&\hbox{(b)}& \langle 0|a_i^+ =0, \q i=1,\l,n,\label{2.7} \\
&\hbox{(c)}& ((a_1^+)^{m_1}(a_2^+)^{m_2}\c(a_n^+)^{m_n}|0\ra,
     (a_1^+)^{l_1}(a_2^+)^{l_2}\c(a_n^+)^{l_n}|0\ra)= \nn\\
&& \qquad \langle 0|(a_n^-)^{m_n}\c(a_2^-)^{m_2}(a_1^-)^{m_1}
     (a_1^+)^{l_1}(a_2^+)^{l_2}\c(a_n^+)^{l_n}|0\ra).\nn
\eea
With respect to this form any two different vectors~(\ref{2.5}) 
are orthogonal.
All vectors
\beq
|p;l_1,\l,l_n\ra=\sqrt{(p-\sum_{j=1}^n l_j )!\over p!}
{(a_1^+)^{l_1}\c(a_n^+)^{l_n}\over{\sqrt{l_1!l_2!\c
l_n!}}}|0\ra, \quad l_1+l_2+\c+l_n\le p \label{2.9}
\eeq
constitute an orthonormal basis in $W(p,n)$, i.e.~$(~,~)$ is a
scalar product. Moreover
the Hermitian conjugate to $a_i^-$ is $a_i^+$,
$(a_i^-)^*=a_i^+$, which is an important physical requirement.

The transformation of the basis~(\ref{2.9}) under the action of the
Jacobson CAO's reads~:
\bea
a_i^+|p;l_1,\l ,l_i,\l,l_n\ra &=&
  \sqrt{(l_i+1)(p-\sum_{j=1}^n l_j  )}~
  |p;l_1\l,l_{i-1},l_i+1,l_{i+1}\l,l_n\ra, \label{2.10a} \\
a_i^-|p;l_1,\l,l_i,\l,l_n\ra &=&
  \sqrt{l_i(p-\sum_{j=1}^n l_j +1  )}~
  |p;l_1\l,l_{i-1},l_i-1,l_{i+1}\l,l_n\ra. \label{2.10b}
\eea
For further use we extend $W(p,n)$ to an irreducible $gl(n+1)$ module,
setting (below and throughout $N_i=e_{ii}$, $i=0,1,\l,n$)
\beq
N_{0} |p;l_1,l_2,\l,l_n\ra=(p-\sum_{i=1}^n
l_i)|p;l_1,l_2,\l,l_n\ra. \label{2.11}
\eeq
Then
\beq
N_{i}|p;l_1,l_2,\l,l_n\ra=l_i|p;l_1,l_2,\l,l_n\ra, \q i=1,\l,n. 
\label{2.12}
\eeq

The basis vectors $|p;l_1,\l,l_n\ra$ in $W(p,n)$
are in one to one correspondence with all distinct
$n$-tuples $(l_1,\l,l_n)$ with integer non-negative entries   
$l_1,\l,l_n$ such that $l_1+\c+l_n \le p$. Based on this we often 
write $(l_1,\l,l_n)$ instead of $|p;l_1,\l,l_n\ra$.

In the present paper we will study some macroscopic properties of
$A$-statistics for a Hamiltonian which is a simple sum 
\beq
H= \sum_{i=1}^n \e_i N_{i}.    \label{2.13}
\eeq
This Hamiltonian can also be written entirely via creation and  
annihilation operators~:
\beq
H={1\over{n+1}}\sum_{i=1}^n \e_i\Big(p+n[a_i^+,a_i^-]- \sum_{k\ne
i=1}^n [a_k^+,a_k^-]\Big).
\label{2.13a}
\eeq

Clearly, $H$ is an element from the Cartan subalgebra of $gl(n+1)$.
Since $H|0\ra=0$, the energy of the vacuum is
zero. The commutation relations of $H$ with the CAO's read~:
\beq
[H,a_i^\pm]=\pm \e_i a_i^\pm. \label{2.14}
\eeq
If $|E\ra$ is a state with energy $E$, then 
\beq
H a_i^\pm|E\ra=(E\pm\e_i)a_i^\pm|E\ra  \label{2.15}
\eeq
and therefore each $a_i^+$ (resp.\ $a_i^-$) can be interpreted 
as an operator creating (resp.\ annihilating)
a particle (quasiparticle, excitation) on orbital~$i$
(with energy~$\e_i$).
Since
\beq
H|p;l_1,l_2,\l,l_n\ra=
(\e_1l_1+\e_2l_2+\c+\e_nl_n)|p;l_1,l_2,\l,l_n\ra, \label{2.17}
\eeq
$|p;l_1,l_2,\l,l_n\ra$ is interpreted as
a state with $l_1$ particles on the first orbital, 
$l_2$ particles on the second orbital and so on,
$l_n$ particles on the last orbital. 

The restriction~(\ref{2.6}) expresses the {\it Pauli principle}
of $A$-statistics in $W(p,n)$. It says that the system can accommodate 
up to $p$, but no more than $p$ particles. For this reason 
$A$-statistics falls into the class of exclusion statistics
in the broad sense~:
the number of the allowed particles to be accommodated 
on a certain orbital depends on the number of the particles that 
have already been accommodated in the system. 
This is perhaps the simplest form of a statistical interaction~:
the Hamiltonian~(\ref{2.13}) has the form of a ``free'' Hamiltonian
and the interaction is introduced via a change of statistics.
It will be interesting to find out whether one can obtain the same
results adding to the Hamiltonian~(\ref{2.13}) an interaction term
and changing the statistics to Bose statistics. It is known that
a similar phenomenon can take place in quantum mechanics~\cite{Poly}.

\section{The grand partition function}
\setcounter{equation}{0}

Here we shall study some  macroscopic properties of
$A$-statistics.  For our considerations it is irrelevant whether
the different orbitals correspond to different particles, to different
energy levels of particles of the same kind or to different internal
states of the particles. The only assumption is that they satisfy the
Pauli principle for $A$-statistics.

As usually, we assume that the system is in a thermal and diffusive
contact and in a thermal and diffusive equilibrium with a much
bigger reservoir. Denote by $\tau$ its (fundamental) temperature
and let $\mu_i$ be the chemical potential for the particles on
orbital~$i$. 

The general principles (and approximations)
of statistical thermodynamics assert that
the probability
$\P(p,n;r)$ for the system to
be in a (quantum) state $r=(l_1,\l,l_n)$ with the number of 
particles $N_r=l_1+\c+l_n$ and energy $E_r=l_1\e_1+\c+l_n\e_n$
is given by the expression~:
\beq
\P(p,n;r)={\exp\Big(
{\sum_{i=1}^n \tau^{-1}(\mu_i -\e_i)l_i}\Big)\over  
Z(p,n)}. \label{3.1}
\eeq
The numerator in~(\ref{3.1}) is the Gibbs factor of the system in the 
state $r=(l_1,\l,l_n)$ and $Z(p,n)$ is the grand partition function 
(GPF), namely the sum of the Gibbs factors with respect to all states
$(l_1,\l,l_n)$ of the system, i.e.~over all possible non-negative 
integers $l_1,\l,l_n$ such that $0\le l_1+\c+l_n \le p$~:
\beq
Z(p,n)=
\sum_{0\le l_1+\c+l_n \le p} (\exp({{\mu_1-\e_1}\over \tau}))^{l_1}
(\exp({{\mu_2-\e_2}\over \tau}))^{l_2}\c 
(\exp({{\mu_n-\e_n}\over \tau}))^{l_n}.   \label{3.3}
\eeq
In terms of the notation
\beq
x_i=\exp\Big({{\mu_i-\e_i}\over \tau}\Big),\quad i=1,\l,n, \label{3.4}
\eeq
we rewrite~(\ref{3.3}) as follows,
\beq
Z(p,n)=\sum_{0\le l_1+\c+l_n \le p}
x_1^{l_1}x_2^{l_2}\c x_n^{l_n}= \sum_{k=0}^p \sum_{l_1+\c+l_n=k}
x_1^{l_1}x_2^{l_2}\c x_n^{l_n}. \label{3.5}
\eeq

In the general setting, which we consider so far,
it is appropriate to introduce
the complete symmetric functions
$h_k(x_1,\l,x_n)$, $k=0,1,\l$, which play an important 
role in the theory of symmetric functions~\cite{Macdonald}.
The $k$-th complete symmetric function $h_k(x_1,\l,x_n)$
is the sum of all distinct monomials of total degree~$k$ of the
variables $x_1,x_2,\l,x_n$~:
\beq
h_k(x_1,\l,x_n)=\sum_{l_1+\c+l_n=k}
x_1^{l_1}x_2^{l_2}\c x_n^{l_n}. \label{3.6}
\eeq
For example, $h_0(x_1,x_2,x_3)=1$, $h_1(x_1,x_2,x_3)=x_1+x_2+x_3$,
\[
h_2(x_1,x_2,x_3)=x_1^2+x_2^2+x_3^2+x_1x_2+x_1x_3+x_2x_3.
\]

In terms of $h_k(x_1,\l,x_n)$, the GPF $Z(p,n)$ reads~:
\beq
Z(p,n)= \sum_{k=0}^p h_k(x_1,\l,x_n). \label{3.7}
\eeq
Clearly, $h_k(x_1,\l,x_n)/Z(p,n)$ yields the
probability for the system to contain $k$ particles.

In order to evaluate the sum~(\ref{3.7}) we use the following
generating function~\cite[(I.2.5)]{Macdonald},
\beq
\sum_{k=0}^\infty h_k(x_1,\l,x_n)t^k
={1\over{(1-x_1t)(1-x_2t)\l(1-x_nt)}}. \label{3.8}
\eeq
Now compute
\beas
&&\sum_{p=0}^\infty Z(p,n)t^p=
\sum_{p=0}^\infty \Big(\sum_{k=0}^p h_k(x_1,\l,x_n)\Big)t^p\\
&&=\sum_{k=0}^\infty \sum_{p=k}^\infty h_k(x_1,\l,x_n)t^p=
\sum_{k=0}^\infty \sum_{r=0}^\infty h_k(x_1,\l,x_n)t^{p+r} \\
&&=\Big(\sum_{k=0}^\infty h_k(x_1,\l,x_n)t^k\Big)
\Big(\sum_{r=0}^\infty t^r\Big)\\
&&={1\over{(1-x_1t)(1-x_2t)\l(1-x_nt)}}
{1\over {1-t}}=\sum_{p=0}^\infty h_p(x_1,\l,x_n,1)t^p.
\eeas
Hence
\beq
\sum_{k=0}^p h_k(x_1,\l,x_n)=h_p(x_1,\l,x_n,1)
=h_p(x_1,\l,x_{i-1},1,x_{i},\l,x_n). \label{3.8a}
\eeq
We have written the last term in the rhs of~(\ref{3.8a}) for further
use. It follows from the property
that $h_p$ is symmetric with respect to its arguments and
therefore these arguments can be reordered in an arbitrary way.

Applying~(\ref{3.8a}) to~(\ref{3.7}) we obtain
\beq
Z(p,n)=
\sum_{k=0}^p h_k(x_1,\l,x_n)=h_p(x_1,\l,x_n,1). \label{3.9}
\eeq

Using the GPF~(\ref{3.9}) one can determine various other 
thermodynamical quantities
and in particular the average number of 
particles in the system. 

According to~(\ref{3.1}) the probability $\P(p,n;l_1,\l,l_n)$
for the system to be in the state $r=(l_1,\l,l_n)$ with 
$N_r=l_1+\c+l_n$ particles reads (in terms of the variables
$x_i$)~:
\beq
\P(p,n;l_1,\l,l_n)=
{{x_1^{l_1}x_2^{l_2}\c x_n^{l_n}}\over Z(p,n)}. \label{3.10}
\eeq
Then the average number of particles in the system is
\beas
{\bar N}(p,n)&=&\sum_{0\le l_1+\c+l_n \le p}(l_1+\c+l_n)
 \P(p,n;l_1,\l,l_n)\\
&=&\sum_{0\le l_1+\c+l_n \le p}(l_1+\c+l_n)
{{x_1^{l_1}\c x_n^{l_n}}\over Z(p,n)},
\eeas
which can also be written as
\beq
{\bar N}(p,n)=\sum_{k=1}^n x_k \partial_{x_k} \ln Z(p,n)=
\sum_{k=1}^n \tau \partial_{\mu_k} \ln Z(p,n).
\label{3.11}
\eeq

Since
\beq
\sum_{0\le l_1+\c+l_n \le p}(l_1+\c+l_n) x_1^{l_1}\c x_n^{l_n}
=\sum_{k=0}^p k \sum_{l_1+\c+l_n=k}x_1^{l_1}\c x_n^{l_n}=
\sum_{k=0}^p k h_k(x_1,\l,x_n),
\label{3.11a}
\eeq
${\bar N(p,n)}$ can also be expressed via the complete symmetric functions,
\beq
{\bar N}(p,n)={\sum_{k=0}^p k h_k(x_1,\l,x_n)
\over{h_p(x_1,\l,x_n,1)}}.    \label{3.13}
\eeq
In order to further simplify~(\ref{3.13}), 
note that according to~(\ref{3.8a})
\[
h_{p-1}(x_1,\l,x_n,x_{n+1},1)=\sum_{k=0}^{p-1}h_k(x_1,\l,x_n,x_{n+1}).
\]
Hence, setting $x_{n+1}=1$ and using again~(\ref{3.8a}) one has
\[
h_{p-1}(x_1,\l,x_n,1,1)=\sum_{k=0}^{p-1}h_k(x_1,\l,x_n,1)
=\sum_{k=0}^{p-1}\sum_{q=0}^{k} h_q,
\]
where here and below $h_q\equiv h_q(x_1,\l,x_n)$.
Therefore
\beq
h_{p-1}(x_1,\l,x_n,1,1)=\sum_{k=0}^{p-1}\sum_{q=0}^{k} h_q
=\sum_{q=0}^{p-1}\sum_{k=q}^{p-1}h_q=\sum_{q=0}^{p-1}(p-q)h_q. 
\label{3.14}
\eeq
From here and~(\ref{3.8a}) we deduce
\beq
ph_p(x_1,\l,x_n,1)-h_{p-1}(x_1,\l,x_n,1,1)=\sum_{k=0}^p p h_k 
-\sum_{k=0}^{p-1} (p-k)h_k = \sum_{k=0}^p k h_k(x_1,\l,x_n).
\label{3.15}
\eeq
Combining~(\ref{3.13}) with~(\ref{3.15}) we finally obtain
\beq
{\bar N}(p,n)=p-{h_{p-1}(x_1,\l,x_n,1,1)\over h_{p}(x_1,\l,x_n,1)}.
\label{3.16}
\eeq
As expected, the average number of particles accommodated in
the system cannot exceed~$p$.

Similarly for the average energy ${\bar E}(p,n)$ of the system one has
\[
{\bar E}(p,n)=\sum_{0\le l_1+\c+l_n \le p}(\e_1l_1+\c+\e_nl_n)
{{x_1^{l_1}x_2^{l_2}\c x_n^{l_n}}\over Z(p,n)},
\]
and therefore
\beq
{\bar E}(p,n)
=\sum_{i=1}^n \e_i x_i\partial_{x_i}\ln h_p(x_1,\l,x_n,1)
=\sum_{i=1}^n \e_i x_i\partial_{x_i}\ln Z(p,n).
\label{3.17}
\eeq

Let us determine the equilibrium distribution of the
particles on an arbitrarily chosen orbital~$i$. 
According to~(\ref{3.10}), $\P(p,n;l_1,\l,l_n)$
yields the probability for the system to be in the state
$(l_1,\l,l_n)$, which means
that $l_1$ particles are accommodated 
on the first orbital,
$l_2$ particles on the second, and so on. Therefore
the probability $\P(p,n;l_i)$ that $l_i$
particles are accommodated on the $i$-th orbital is
\beq
\P(p,n;l_i)=\sum_{0\le l_1+\c+l_{i-1}+l_{i+1}+\c+l_{n}\le p-l_i}
{{x_1^{l_1}x_2^{l_2}\c x_{n}^{l_{n}}}\over Z(p,n)}.
\label{3.19}
\eeq
For the average number of
particles ${\bar l}_i$ on the $i$-th orbital we have
\beas
{\bar l}_i&=&\sum_{l_i=0}^p l_i \P(p,n;l_i)=
\sum_{0\le l_1+\c+l_{n}\le p}l_i
{{x_1^{l_1}x_2^{l_2}\c x_{n}^{l_{n}}}\over Z(p,n)}\\
&=&{1\over Z(p,n)}x_i\partial_{x_i}
\sum_{0\le l_1+\c+l_{n}\le p}
{x_1^{l_1}x_2^{l_2}\c x_{n}^{l_{n}}}.
\eeas
Hence
\beq
{\bar l}_i
=x_i\partial_{x_i}\ln h_p(x_1,\l,x_n,1)
= x_i\partial_{x_i}\ln Z(p,n)=\tau \partial_{\mu_i}\ln Z(p,n),
\quad i=1,\l,n.     
\label{3.20}
\eeq
It follows that the average number of particles
$N_A(p,n)$ on, say,  the first $s$ orbitals is
\beq
N_A(p,n)=\sum_{i=1}^s {\bar l}_i=\sum_{i=1}^s  x_i\partial_{x_i}\ln Z(p,n)
=\sum_{i=1}^s \tau \partial_{\mu_i}\ln Z(p,n).
\label{3.20+}
\eeq

Evidently, the average energy ${\bar E}_i$ 
of the particles on the $i$-th orbital is~:
\beq
{\bar E}_i=
\e_i x_i\partial_{x_i}\ln Z(p,n)=\e_i\tau \partial_{\mu_i}\ln Z(p,n),
\quad i=1,\l,n.     
\label{3.20a}
\eeq

Let us note that the expression for the probability~(\ref{3.19}) 
can be also written in a more compact form~:
\beas
\P(p,n;l_i)&=&{x_i^{l_i}\over Z(p,n)} \ 
\sum_{0\le l_1+\c+l_{i-1}+l_{i+1}+\c+l_{n}\le p-l_i}
{x_1^{l_1}\c x_{i-1}^{l_{i-1}}x_{i+1}^{l_{i+1}}\c x_{n}^{l_{n}}}\\
&=&{x_i^{l_i}\over Z(p,n)}\sum_{k=0}^{p-l_i} \
\sum_{l_1+\c+l_{i-1}+l_{i+1}+\c+l_{n}=k} 
{x_1^{l_1}\c x_{i-1}^{l_{i-1}}x_{i+1}^{l_{i+1}}\c x_{n}^{l_{n}}}\\
&=&{x_i^{l_i}\over Z(p,n)}\sum_{k=0}^{p-l_i}
h_k(x_1,\l,x_{i-1},x_{i+1},\l,x_n).
\eeas
Applying~(\ref{3.8a}) to the rhs, we obtain the required expression
for the probability to have $l_i$ particles accommodated
on the $i$-th orbital~:
\beq
\P(p,n;l_i)={h_{p-l_i}(x_1,\l, x_{i-1},1,x_{i+1}\l x_{n})x_i^{l_i}\over Z(p,n)}
={h_{p-l_i}(x_1,\l, x_{i-1},1,x_{i+1}\l x_{n})x_i^{l_i}
\over h_{p}(x_1,\l,x_{n},1)}.
\label{3.21}
\eeq

Some other thermodynamical functions can be determined too. For instance,
from the general expression for the entropy
\beq
S(p,n)={{\bar{E}(p,n)-\sum_{i=1}^n \mu_i \bar{l}_i}\over T} 
+ k_B \ln Z(p,n)  
\label{3.21a}
\eeq
and~(\ref{3.17}) there comes~:
\beq
S(p,n)
={k_B\over \tau} \sum_{i=1}^n (\e_i-\mu_i) \bar{l}_i+ k_B \ln Z(n,p)=
{k_B\over \tau} \sum_{i=1}^n (\e_i-\mu_i) x_i \partial_{x_i}\ln Z(p,n)
+ k_B \ln Z(n,p), 
\label{3.21b}
\eeq
which can also be written as
\beq
S(p,n)=k_B \Big(\tau \partial_{\tau} +1 \Big)\ln Z(p,n)
=k_B \partial_{\tau} \tau \ln Z(p,n),      
\label{3.21c}
\eeq
or equivalently
\beq
S(p,n)=k_B \tau \partial_{\tau} \ln Z(p,n)
- {k_B\over \tau}\Omega, 
\label{3.21d}
\eeq
where 
\beq
\Omega=-\tau \ln Z(p,n) 
\label{3.21e}
\eeq
is the thermodynamical potential, another relevant thermodynamical 
function
(in order to be consistent with the notation used so far we have
replaced in~(\ref{3.21b})-(\ref{3.21e}) the Kelvin temperature~$T$ 
with the fundamental temperature $\tau=k_B T$, $k_B$ being the 
Boltzmann constant).

Before proceeding further with some particular cases of the 
Hamiltonian~(\ref{2.13}) we make a small deviation
in order to draw a parallel between $A$-statistics and Bose statistics. 
To this end introduce new creation and annihilation operators
\beq
B(p)_i^\pm = {a_i^\pm\over \sqrt{p}}, \q i=1,\l,n, \q p\in \N,
\label{3.29+1}
\eeq
in $W(p,n)$. It is easy to verify that for large values of $p$ these
operators satisfy ``almost Bose" commutations relations~\cite{PV00}~:
\bea
&& [B(p)_i^+,B(p)_j^+]=[B(p)_i^-,B(p)_j^-]=0,\q 
\hbox{exact commutators}, \label{3.29+2a} \\
&& [B(p)_i^-,B(p)_j^+]\simeq
\delta_{ij}, \q\hbox{ if } l_1+l_2+\c+l_n \ll p. \label{3.29+2b}
\eea
Therefore the representations of $B(p)_i^\pm$ in the Fock
spaces $W(p,n)$ with large values of~$p$, restricted to states
with a small number $l_1+l_2+\c+l_n \ll p$ of accommodated
particles provide good approximations to Bose creation and
annihilation operators (in finite-dimensional spaces).
For this reason the operators $B(p)_i^\pm$
are said to be quasi-Bose creation and annihilation operators  
(of order~$p$). In the limit $p\rightarrow \infty$ these operators 
become indeed Bose operators~\cite{PV00}.
Therefore, parallel to quon statistics (see~\cite{Greenberg2}
and references therein), $A$-statistics (for large values of $p$)
can be considered as a theory allowing small violations of canonical
quantum statistics in nonrelativistic quantum field theory.

Coming back to the macroscopic considerations, we observe that
for $t=1$ the rhs of~(\ref{3.8}) reduces to the Bose GPF $Z_{Bose}(n)$
of a system with $n$ orbitals, which are filled independently of each
other~:
\beq
\sum_{k=0}^\infty h_k(x_1,\l,x_n)
={1\over{(1-x_1)(1-x_2)\c(1-x_n)}}
=Z_{Bose}(n). \label{3.29+3}
\eeq
Therefore, see~(\ref{3.9}), 
\beq
Z_{Bose}(n)-Z(p,n)=
\sum_{k=p+1}^\infty h_k(x_1,\l,x_n). \label{3.29+4}
\eeq
For sufficiently large values of~$p$ the rhs of~(\ref{3.29+4}),
which is always positive, can 
be made smaller than any positive number and therefore can be neglected. 
This is another confirmation (now from a macroscopic point of view) that
$A$-statistics reduces to Bose statistics as the order
of statistics~$p$ becomes large. 
An analogue of equation~(\ref{3.29+4}) for $q$-statistics is
also available~\cite[eq.~(5.14)]{Chaichian}.

In the next sections we shall consider some examples, the first
one with all energies equal to each other.

\section{The most degenerate case}
\setcounter{equation}{0}

Here we consider an ensemble of particles with a Hamiltonian  
\beq
H=\e \sum_{i=1}^n N_i, \label{3.30}
\eeq
i.e.\ all orbitals have the same energy,
and additionally we assume that they all have the same chemical potential,
i.e.,
\beq
\e_1=\e_2=\l=\e_n=\e, \quad
\mu_1=\l=\mu_n=\mu \q\Longrightarrow \q x_1=x_2=\l=x_n=x. 
\label{3.30+}
\eeq
In this case the orbitals label internal degrees of freedom of the
particles (spin, color, flavor) or, as more particular examples, 
the local orbitals of any multi-band Hubbard model or $SU(N)$ 
Heisenberg chain.

Most of the thermodynamical functions follow directly from the
results of the previous section after the specialization~(\ref{3.30+}),
but they can be written in a more explicit form. 
To this end one has to take into account that the number of terms
in the rhs of~(\ref{3.6}) is $(k+n-1)!/k!(n-1)!$. Therefore
\beq
h_k(\underbrace{x,\ldots,x}_{\hbox{$n$ times}})=
{k+n-1 \choose k} x^k. 
\label{3.30a}
\eeq
Then equation~(\ref{3.9}) yields
\beq
Z(p,n)=h_p(\underbrace{x,\ldots,x}_{\hbox{$n$ times}},1)=
\sum_{k=0}^p {k+n-1 \choose k} x^k. 
\label{3.31}
\eeq
This sum can be rewritten as
\bea
Z(n,p)&=&\sum_{k=0}^\infty {k+n-1 \choose k} x^k -
\sum_{k=p+1}^\infty {k+n-1 \choose k} x^k \nn\\
&=& {1\over (1-x)^n} - {n+p\choose p+1} x^{p+1}\;{}_2F_1\left({1,n+p+1 \atop
p+2};x\right), \label{3.32}
\eea
where ${}_2F_1$ is the classical hypergeometric function~\cite{Slater}.
Compared to~(\ref{3.31}), the expression in~(\ref{3.32}) looks at first sight
a more complicated way of rewriting $Z(n,p)$. Note, however,
that the first term in the rhs of~(\ref{3.32}) is the Bose GPF
\beq
Z(n)_{\hbox{Bose}} = \sum_{k=0}^\infty {k+n-1 \choose k} x^k =
{1\over (1-x)^n}. \label{3.32a}
\eeq
Therefore, the second term
is responsible for the difference between Bose and  
$A$-statistics. It carries, so to speak, the statistical
interaction between the particles.

Using Euler's transformation formula for hypergeometric 
functions~\cite[(1,3,15)]{Slater}, i.e.
\beq
{}_2F_1\left({a,b \atop c};x\right) = (1-x)^{c-a-b}
{}_2F_1\left({c-a,c-b \atop c};x\right),   \label{3.32b}
\eeq
(\ref{3.32}) can also be rewritten as
\beq
Z(n,p)={1\over (1-x)^n} \left( 1- {n+p\choose p+1} x^{p+1}\;
{}_2F_1\left({p+1,1-n \atop p+2};x\right)\right). \label{3.32c}
\eeq
The hypergeometric series appearing in~(\ref{3.32c}) has the advantage
that it is a terminating series (consisting of~$n$ terms), since
one of its numerator parameters, $1-n$, is a negative integer.
More explicitly, we can rewrite~(\ref{3.32c}) as
\beq
Z(n,p)={1\over (1-x)^n} \left( 1- {(n+p)!\over p!}
\sum_{k=0}^{n-1} (-1)^k {x^{p+k+1} \over (p+k+1) k!(n-k-1)!} 
\right). 
\label{3.32d}
\eeq
Equation~(\ref{3.31}) is convenient to deal with in those cases
that the order of statistics~$p$ is a small number (and any 
number of orbitals~$n$).
On the contrary, the expression~(\ref{3.32d}) is more
appropriate for a relatively small number of orbitals (and any
order of statistics~$p$).

In the case of only one orbital, i.e.~for the $sl(2)$ GPF, 
equation~(\ref{3.31}) yields
\beq
Z(p,1)= \sum_{k=0}^p x^k= {{1-x^{p+1}}\over{1-x}}. \label{3.33}
\eeq
For the $sl(3)$ GPF, the expression is
\beq
Z(p,2)=\sum_{k=0}^p(k+1) x^k={1\over{(1-x)^2}}
+{px+x-p+2\over{(1-x)^2}}x^{p+1},
\label{3.34}
\eeq
and it can be related to the $Z(p,1)$ partition function by
\beq
Z(p,2)=\Big({1\over{1!}}{\partial\over{\partial x}}x\Big)Z(p,1)
= \Big({1\over{1!}}{\partial\over{\partial x}}x\Big){{1-x^{p+1}}\over{1-x}}.
\label{3.35}
\eeq
This result can be further generalized. The GPF of $sl(n+1)$ for
any $n$ can be related to the GPF of $sl(2)$~:
\beq
Z(p,n)={1\over{(n-1)!}}\ {\partial^{n-1}\over{\partial x^{n-1}}} \
x^{n-1} \ Z(p,1), \label{3.36}
\eeq
or equivalently
\beq
Z(p,n)={1\over{(n-1)!}}\ {\partial^{n-1}\over{\partial x^{n-1}}} 
\ x^{n-1} \
\sum_{k=0}^p \ x^k =
{1\over{(n-1)!}}\ {\partial^{n-1}\over{\partial x^{n-1}}} \
x^{n-1} \ {{1-x^{p+1}}\over{1-x}}. 
\label{3.37}
\eeq

Clearly, after the specialization~(\ref{3.30+}) the expression~(\ref{3.11})
for the average number of particles reads
\beq
\bar{N}(p,n)= x\partial_x\ln Z(n,p) \label{3.38}
\eeq
and $\bar{E}(p,n)=\e \bar{N}(p,n)$.
Another expression follows from~(\ref{3.13}), (\ref{3.30a}) 
and~(\ref{3.31})~: 
\beq
{\bar N}(p,n)=
{\sum_{k=0}^p k {k+n-1 \choose k} x^k \over 
\sum_{k=0}^p {k+n-1 \choose k} x^k}.    \label{3.39a}
\eeq
Using the definitions of hypergeometric functions, (\ref{3.39a})
can be rewritten as
\beq
\bar N(p,n)= {{nx\over (1-x)^{n+1}} - (p+1){n+p\choose p+1} 
x^{p+1}\;{}_2F_1\left({1,n+p+1 \atop p+1};x\right)
\over {1\over (1-x)^n} - {n+p\choose p+1} 
x^{p+1}\;{}_2F_1\left({1,n+p+1 \atop p+2};x\right) }. \label{3.39b}
\eeq
Applying Euler's transformation to each of the ${}_2F_1$ functions
yields an expression with {\it terminating} hypergeometric
series in the numerator and denominator~:
\beq
\bar N(p,n)={ x \left( n- 
{(p+1)}{n+p\choose p+1} x^{p}\;
{}_2F_1\left({p,-n \atop p+1};x\right)\right)
\over {(1-x)} \left( 1- {n+p\choose p+1} x^{p+1}\;
{}_2F_1\left({p+1,1-n \atop p+2};x\right)\right)}.  \label{3.39c}
\eeq
So we find
\beq
\bar N(p,n)= {nx\over 1-x} \left( { p!- (n+p)!
\sum_{k=0}^{n} (-1)^k {x^{p+k} \over (p+k) k!(n-k)!}
\over p!- (n+p)!
\sum_{k=0}^{n-1} (-1)^k {x^{p+k+1} \over (p+k+1) k!(n-k-1)!}} \right). 
\label{3.39d}
\eeq
The last expression for ${\bar N}(p,n)$ is more appropriate to work with 
for small values of $n$, whereas~(\ref{3.39a}) is more suitable 
for small values of $p$.

{}From~(\ref{3.21}) and~(\ref{3.30a}) we can also compute the probabiliy 
$\P(p,n;l_i)$ that $l_i$
particles are accomodated on the $i$-th orbital~:
\beq
\P(p,n;l_i)= {1\over Z(p,n)}
\sum_{k=0}^{p-l_i}{k+n-2\choose k} x^{k+l_i}.
\label{3.39e}
\eeq
Then the average number of particles accommodated on 
the $i$-th orbital is
\beq
 {\bar l}_i= 
\sum_{l=0}^p l\ \P(p,n;l)={1\over Z(p,n)}
\sum_{l=0}^p
\sum_{k=0}^{p-l}l{k+n-2\choose k} 
x^{k+l}. 
\label{3.39f}
\eeq
As it should be, the result does not depend on the number~$i$ of
the orbital~:
${\bar l}_1=\c ={\bar l}_i=\c={\bar l}_n\equiv {\bar l}$. 

Using the binomial identity 
\beq
\sum_{l=0}^r l \ {r-l+n-2\choose r-l}={r\over n}{n+r-1\choose r}.
\label{3.39g} 
\eeq
one verifies that the consistency condition ${\bar N}(p,n)=n{\bar l}$ holds
too. Hence
\beq
{\bar l}_i={ x \left( n- 
{(p+1)}{n+p\choose p+1} x^{p}\;
{}_2F_1\left({p,-n \atop p+1};x\right)\right)
\over {n(1-x)} \left( 1- {n+p\choose p+1} x^{p+1}\;
{}_2F_1\left({p+1,1-n \atop p+2};x\right)\right)},~~i=1,\l,n.
\label{3.39h}
\eeq
Other thermodynamical functions follow straightforwardly.
Equation~(\ref{3.21b}) for the entropy reduces to~:
\beq
S(p,n)={k_B\over \tau}(\e-\mu){\bar N}(p,n)-
{k_B\over \tau}\Omega, 
\label{3.39i}
\eeq
where $\Omega=-\tau \ln Z(p,n)$ is the thermodynamical 
potential~(\ref{3.21d}).

Let us consider in some more detail the dependence on the
energy of the average number of particles in
the system $\bar{N}(p,n)$, i.e.\ the distribution function.
As an energy variable we take
\beq
y={{\e-\mu}\over{\tau}} \q\rightarrow\q x=e^{-y}, \label{3.42}
\eeq
namely the energy in units of $\tau$.
We will consider two extreme cases.

\begin{itemize}
\item
$n=1$ and any $p$~:
\beq
\bar{N}(p,1)={1\over{e^y-1}}-{(p+1)\over{e^{(p+1)y}-1}}.
\label{3.43}
\eeq
Note that at $p=1$ one obtains the Fermi-Dirac distribution,
\beq
\bar{N}(1,1)={1\over{e^{({\e-\mu)\over \tau}}+1}}.
\label{3.44}
\eeq
In Figure~1 we plot $\bar{N}(p,1)$ for $p=1,2,3,4,6,8,12,16,\infty$.

\begin{figure}[htb]
\caption{Graph of $\bar{N}(p,1)$ for $p=1,2,3,4,6,8,12,16,\infty$}
\[
\includegraphics{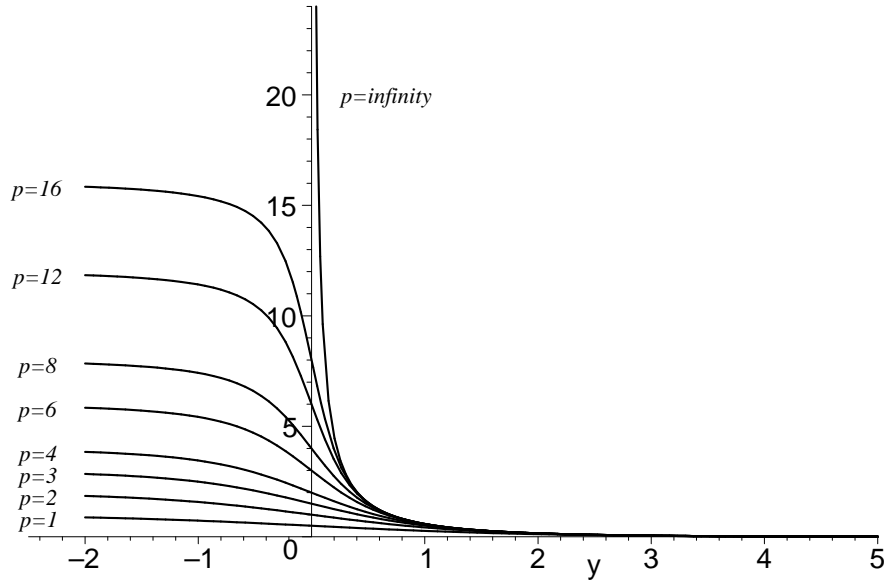}
\]
\end{figure}

The lowest curve $(p=1)$ yields the Fermi-Dirac distribution.
Increasing $p$ from $1$ to $\infty$ one ``deforms" it into 
the Bose-Einstein distribution $(p=\infty)$ 
\beq
\bar{N}(p,\infty)={1\over{e^{({\e-\mu)\over \tau}}-1}}.
\label{3.45}
\eeq

\item $p=1$ and any $n$~:
\beq
{\bar N}(1,n)={n\over{e^{({\e-\mu)\over \tau}}}+n}. 
\label{3.46}
\eeq
${\bar N}(1,n)$ is always smaller then 1, 
so the system can accommodate at most one particle. 
As an example we plot the distribution functions for a
system with $n=1,2,4,8,16, 32, 64, 128$ orbitals (Figure~2).

\begin{figure}[htb]
\caption{Graph of $\bar N(1,n)$ for $n=1,2,4,8,16, 32, 64, 128$}
\[
\includegraphics{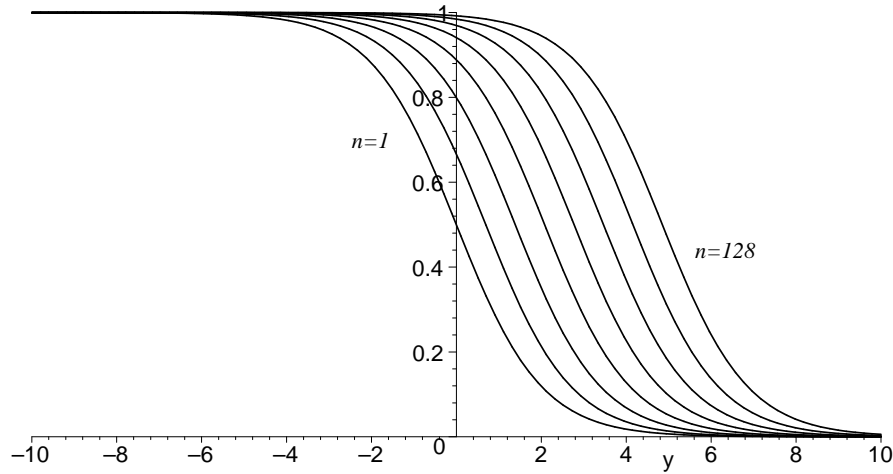}
\]
\end{figure}

The first curve $n=1$ (from the left) corresponds to the
Fermi-Dirac distribution function. With the increase of the
number of orbitals the average occupation number of the system
increases for fixed $y$. In particular for $\e=\mu$ we have that
${\bar N}(1,n)={n\over{n+1}}$. All curves are ``Fermi-like" but the
half-filling is shifted to the right, at $y=\ln n$.

It should be noted that the curves in Figure~2 give 
the average number of particles in the system, not on a certain
orbital. 

The particles described above ($p=1,~n>1$) are called hard-core bosons.
They appear naturally in various models of condensed matter
physics and nuclear physics (for more discussions and references
see~\cite{PV00}).
\end{itemize}

\section{Equidistant energy levels}
\setcounter{equation}{0}

Let us now consider the Hamiltonian~(\ref{2.13})
with equidistant energies $\epsilon_i$.
Denote the gap between the different energy
levels by $\Delta >0$. 
This means that $\epsilon_2=\epsilon_1+\Delta$, 
$\epsilon_3=\epsilon_1+2\Delta$, etc., or
\beq
\epsilon_i=\epsilon_1+(i-1)\Delta, \qquad (i=1,2,\ldots,n).
\label{3.55}
\eeq
Just as in the previous section, we shall assume
that $\mu_1=\mu_2=\cdots=\mu_n=\mu$.
In this setting the different orbitals correspond to different
energy levels.
Following the notation of~(\ref{3.4}), we have
\beq
x_i= \exp \left( {\mu-\epsilon_i\over \tau}\right)
= \exp \left( {\mu-\epsilon_1\over \tau}\right)
\exp \left(- {\Delta\over \tau}\right)^{i-1}
= x q^{i-1}, 
\label{3.56}
\eeq
where we have used the notation
\beq
x=x_1=\exp \left( {\mu-\epsilon_1\over \tau}\right)
\qquad\hbox{ and }\qquad 
q=\exp \left( -{\Delta\over \tau}\right).
\label{3.57}
\eeq

In order to write down the grand partition function, we can 
use~(\ref{3.9}) and the specialization given above~:
\beq
Z(p,n)=\sum_{k=0}^p h_k(x,qx,q^2x,\ldots,q^{n-1}x)
=h_{p}(x,qx,q^2x,\ldots,q^{n-1}x,1).
\label{3.58}
\eeq
The symmetric functions simplify under this specialization. 
To see this, consider their generating function~(\ref{3.8}). 
Since~\cite[p.~26]{Macdonald}
\beq
{1\over (1-xt)(1-qxt)\cdots(1-q^{n-1}xt)} 
= \sum_{k=0}^\infty \left[ n+k-1 \atop k \right] x^k t^k,
\label{3.59}
\eeq
where $\left[ m \atop k \right]$ denotes the $q$-binomial
coefficient or Gaussian polynomial~\cite[p.~26]{Macdonald}~:
\beq
\left[ m \atop k \right] = 
{(1-q^m)(1-q^{m-1})\cdots (1-q^{m-k+1}) \over
(1-q)(1-q^2) \cdots (1-q^k)},
\label{3.60}
\eeq
it follows from~(\ref{3.8}) that
\beq
h_k(x,qx,q^2x,\ldots,q^{n-1}x)= \left[ n+k-1 \atop k \right] x^k.
\label{3.61}
\eeq
Observe that in the limit $q \rightarrow 1$, the $q$-binomial
$\left[ m \atop k \right]$ goes to the ordinary binomial
coefficient ${m\choose k}$.
Using (\ref{3.61}), (\ref{3.9}) implies
\beq
Z(p,n) = \sum_{k=0}^p \left[ n+k-1 \atop k \right] x^k.
\label{3.62}
\eeq
Using the $q$-raising factorials~\cite{GR},
\beq
(a;q)_k = (1-a)(1-qa)\cdots (1-q^{k-1}a),
\label{3.63}
\eeq
and the classical $q$-generalized hypergeometric series,
called basic generalized hypergeometric 
series~\cite{GR,Slater}, this can be rewritten as
\beq
Z(p,n) = \sum_{k=0}^p { (q^n;q)_k \over (q;q)_k} x^k
= {}_2\Phi_1 \left( {q^n, q^{-p} \atop q^{-p} } ; x \right).
\label{3.64}
\eeq

The average number of particles in the system follows from~(\ref{3.13})~:
\beq
\bar N(p,n) = { \sum_{k=0}^p k \left[ n+k-1 \atop k\right] x^k \over
\sum_{k=0}^p \left[ n+k-1 \atop k\right] x^k }
= x\ {\partial\over \partial x} ( \ln Z(p,n) ) \quad
\Big(= \tau\ {\partial\over \partial \mu}( \ln Z(p,n) )\Big).
\label{3.67}
\eeq
This expression cannot be further simplified.

Another quantity that carries relevant
information about the system is the average number
of particles accommodated on a particular orbital.
Let $\bar l_i$ be this average for the $i$-th orbital,
$i=1,2,\l,n$. Following~(\ref{3.20}), we have 
\beq
\bar l_i = {1\over Z(p,n)}x_i \partial_{x_i}(Z(p,n)),
\label{3.68}
\eeq
in which we have to substitute $x_i=q^{i-1}x$. 
This expression can be written in the following more explicit form~:
\beq
\bar l_i = {1\over Z(p,n)}\sum_{r=1}^p (q^{i-1}x)^r \sum_{l=0}^{p-r}
\left[ n+l-1 \atop l \right] x^l.
\label{3.69}
\eeq
The derivation of~(\ref{3.69}), which is not so trivial, is given in
the Appendix.

The main conclusion from~(\ref{3.69}) is that the ``population" 
of the orbitals depends essentially on their level~$i$ 
via $q^{i-1}$, where $q=\exp \left( -{\Delta\over \tau}\right)<1$~:
as~$i$ grows, the average number of particles $\bar l_i$ decreases.
Otherwise said~: the higher the energy level, the lower the average
number of particles. 

If we consider the extreme case $p=1$, where the system contains
only one particle, and any $n$ (the other
extreme case, any $p$ and $n=1$ coincides with the most
degenerate case) then there comes
\beq
\bar N(1,n) = { (1+q+\cdots+q^{n-1}) \over e^{\beta(\e_1-\mu)} 
+ (1+q+\cdots+q^{n-1})},\quad \beta = {1\over \tau}.
\label{3.71}
\eeq
The case $q=1$ ($\Delta=0$) corresponds to the degenerate case. 

For values of $q=\exp(-\Delta/\tau)\ll 1$, i.e., for
large gaps between the energy levels or very low temperature, 
one can neglect all positive powers of $q$ in~(\ref{3.71}). 
What remains is the Fermi-Dirac distribution 
\beq
\bar N(1,n) \approx {1\over {e^{\beta(\e_1-\mu)} +1}}. 
\label{3.72}
\eeq

Continuing with this extreme case (where $p=1$), 
the expression for the average number of particles
on orbital~$i$ reads
\beq
\bar l_i = {q^{i-1} \over \e^{\beta(\e_1-\mu)}+(1+q+\cdots+q^{n-1})}, 
\quad i=1,\l,n.
\label{3.73}
\eeq
For very low temperatures, or big $\Delta$, (\ref{3.73}) reduces to
\beq
\bar l_1 \approx  {1\over {e^{\beta(\e_1-\mu)}+1}} \ \hbox{ and }\
\qquad  \bar l_i \approx 0 \ \hbox{ if }\ i>1. 
\label{3.74}
\eeq
The latter means that if the system contains a particle, it is
``sitting" permanently on the first,
i.e.\ on the lowest energy orbital. This also explains why 
$\bar N(1,n)\approx \bar l_1$. 

The expressions for the entropy $S(p,n)$ and the thermodynamical
potential $\Omega(p,n)$ follow from~(\ref{3.21c})-(\ref{3.21e}) 
and cannot be simplified considerably.

\section{Two species of particles}
\setcounter{equation}{0}

We assume in this section 
that the system under consideration consists of two
species of particles. Those of the first kind~$A$ (resp.\
of the second kind~$B$) have one and the same energy $\e_A$ 
and chemical potential $\mu_A$ (resp.\ $\e_B$ and $\mu_B$).
To be more precise, the Hamiltonian of the system is
\beq
H=\e_A\sum_{i=1}^{m}N_i + \e_B\sum_{i=m+1}^{n}N_i,\quad m={n\over 2}.
\label{3.80}
\eeq
For convenience we consider a system with an even number of 
orbitals~: $n=2m$, $m\in\N$. 
The first $m$ orbitals refer to single particle states of 
kind~$A$, and the remaining $m$ to single particle states of kind~$B$.

The probability for the system to be in a state $r=(l_1,\l,l_{n})$
is given by~(\ref{3.1}), which in this case reads~:
\beq
\P(p,n;r)={x_A^{l_1+\l+l_{m}} x_B^{l_{m+1}+\l+l_{n}}
\over  Z(p,n)}, 
\label{3.81}
\eeq
where 
\beq
x_A=\exp\Big({\mu_A-\e_A\over \tau}\Big),\quad
x_B=\exp\Big({\mu_B-\e_B\over \tau}\Big). 
\label{3.82}
\eeq
In order to write down the grand partition function~(\ref{3.7})
we use the following identity
\beq
h_k(x_1,\l,x_m,\l,x_n)=\sum_{r=0}^k h_r(x_1,\l,x_m)h_{k-r}(x_{m+1},\l,x_n),
\label{3.82a}
\eeq
which can easily be derived from the generating function~(\ref{3.8}).
Then, in view of~(\ref{3.30a})
\[
h_k(\underbrace{x_A,\l,x_A}_{\hbox{$m$ times,}},
    \underbrace{x_B,\l,x_B}_{\hbox{$m$ times}})=
    \sum_{r=0}^k
    {r+m-1 \choose r} {k-r+m-1\choose k-r}x_1^r x_2^{k-r}.
\]
The latter can also be expressed by means of a hypergeometric function~:
\beq
h_k(\underbrace{x_A,\l,x_A}_{\hbox{$m$ times,}},
 \underbrace{x_B,\l,x_B}_{\hbox{$m$ times}})=
{k+m-1\choose k}\
{}_2F_1\left({m,-k \atop 1-m-k};{x_1\over x_2}\right)x_2^k .
\label{3.82b}
\eeq
Hence the GPF~(\ref{3.7}) reduces to the following expression~:
\beq
Z(p,n)=\sum_{k=0}^p \ \sum_{r=0}^k
{r+m-1 \choose r} {k-r+m-1\choose k-r}x_1^r x_2^{k-r},
\label{3.82c}
\eeq
or 
\beq
Z(p,n)=\sum_{k=0}^p 
{k+m-1\choose k}\
{}_2F_1\left({m,-k \atop 1-m-k};{x_1\over x_2}\right)x_2^k.
\label{3.83}
\eeq

An immediate consequence of~(\ref{3.11}) is the expression for the 
average number of particles in the system~:
\beq
 {\bar N}(p,n)=(x_A\partial_{x_A}+x_B\partial_{x_B}) \ln Z(p;n) 
=\tau (\partial_{\mu_A}+
\partial_{\mu_B}) \ln Z(p,n). 
\label{3.84}
\eeq
Using (\ref{3.13}), (\ref{3.82b}) and (\ref{3.83}), 
one can write ${\bar N}(p,n)$
in a more explicit form~:
\beq
{\bar N}(p,n)={{\sum_{k=0}^p k{k+m-1\choose k}
{}_2F_1\left({m,-k \atop 1-m-k};{x_A\over x_B}\right)x_B^k }\over
{\sum_{k=0}^p 
{k+m-1\choose k}
{}_2F_1\left({m,-k \atop 1-s-k};{x_B\over x_B}\right)x_B^k}}.
\label{3.85}
\eeq

{}From (\ref{3.81}) one derives the probability $\P(p,n;M_A,M_B)$ 
for the system
to contain $M_A$ particles of kind $A$ and $M_B$ particles of kind $B$~:
\beq
\P(p,n;M_A,M_B)={1\over Z(p,n)}
{M_A+m-1 \choose M_A} {M_B+m-1 \choose M_B} 
x_A^{M_A} x_B^{M_B}. 
\label{3.87}
\eeq
Consequently
\beq
\P(p,n;M_A) ={1\over Z(p,n)} \sum_{M_B=0}^{p-M_A}
{M_A+m-1 \choose M_A} {M_B+m-1 \choose M_B} 
x_A^{M_A} x_B^{M_B}, 
\label{3.88}
\eeq
yields the probability for the system to accommodate $M_A$ particles of 
kind~$A$. 
Therefore, the thermal average of the particles of kind~$A$ reads~:
\beq
{\bar N}(p,n)_A={1\over Z(p,n)}\sum_{M_A=0}^p M_A 
 \sum_{M_B=0}^{p-M_A}
{M_A+m-1 \choose M_A} {M_B+m-1 \choose M_B} 
x_A^{M_A} x_B^{M_B}. 
\label{3.91}
\eeq
Formulas (\ref{3.88}) and (\ref{3.91}) can be re-expressed
in terms of a hypergeometric function~:
\bea
\P(p,n;M_A)&=&{1\over Z(p,n)}{M_A+m-1\choose M_A}\label{3.89} \\
&\times& \Big({1\over (1-x_B)^{m}}- x_B^{p-M_A+1}{m+p-M_A \choose
m-1}{}_2F_1\left({1,m+p-M_A+1 \atop p-M_A+2};{x_B}\right)
\Big)x_A^{M_A}, \nn
\eea
and
\bea
{\bar N}(p,n)_A&=& {1\over Z(p,n)}\sum_{M_A=0}^p M_A 
{M_A+m-1\choose M_A}\label{3.90}\\
&\times& \Big({1\over(1-x_B)^{m}}- x_B^{p-M_A+1}{m+p-M_A \choose
m-1} {}_2F_1\left({1,m+p-M_A+1 \atop p-M_A+2};{x_B}\right)
\Big)x_A^{M_A}. \nn
\eea
More formally, we can also write
\beq
{\bar N}(p,n)_A=x_A\partial_{x_A} \ln Z(p;n) 
=\tau \partial_{\mu_A}
\ln Z(p,n). 
\label{3.92}
\eeq

The termal averages ${\bar E}(p,n)_A$ and ${\bar E}(p,n)_B$
of the particles of kind~$A$ and~$B$  are evident~:
\beq
{\bar E}(p,n)_A=\e_A {\bar N}(p,n)_A,\qquad
{\bar E}(p,n)_B=\e_B {\bar N}(p,n)_B, 
\label{3.93}
\eeq
and therefore
\beq
{\bar E}(p,n)={\bar E}(p,n)_A+{\bar E}(p,n)_B  
\label{3.94}
\eeq
yields the average energy of the system.
 
{}From (\ref{3.21b})-(\ref{3.21e}) the expression for the entropy
follows~:
\beq
S(p,n)={k_B\over \tau}(\e_A-\mu_A){\bar N}(p,n)_A +
       {k_B\over \tau}(\e_B-\mu_B){\bar N}(p,n)_B -
       {k_B\over \tau}\Omega,  
\label{3.95}
\eeq
where $\Omega=-\tau \ln  Z(p,n)$ is  
the thermodynamical potential.

It is instructive to consider an example in more detail.
Let us fix $n=4$ and also take $p=4$. We shall draw a graph
of the average number of particles $\bar N(p,n)=\bar N(4,4)$
(see equation~(\ref{3.85})), as a function of two energy
variables $y_A$ and $y_B$ associated with the two kinds
of particles of the system, i.e.
\beq
y_A={\epsilon_A-\mu_A\over\tau},\qquad
y_B={\epsilon_B-\mu_B\over\tau}.
\eeq
This graph is given in Figure~3.
\begin{figure}[htb]
\caption{Graph of $\bar N(4,4)$ for $y_A$ and $y_B$ in the range $[-5,5]$.}
\[
\includegraphics{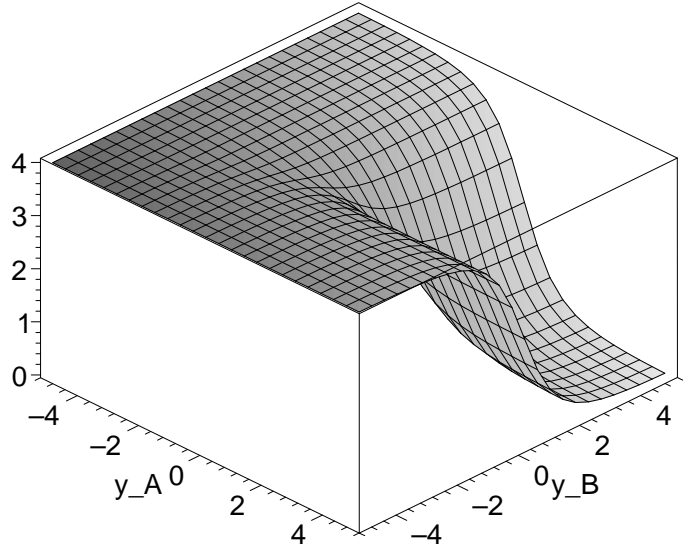}
\]
\end{figure}
Clearly, this graph is symmetric with respect to $y_A$
and $y_B$. 

Let us now also consider, for this same example, the graph
of the average number of particles of kind~$A$, i.e.\
$\bar N_A(4,4)$. The expression follows from~(\ref{3.91}). 
The graph is given in Figure~4.
\begin{figure}[htb]
\caption{Graph of $\bar N_A(4,4)$ for $y_A$ and $y_B$ in the range $[-5,5]$.}
\[
\includegraphics{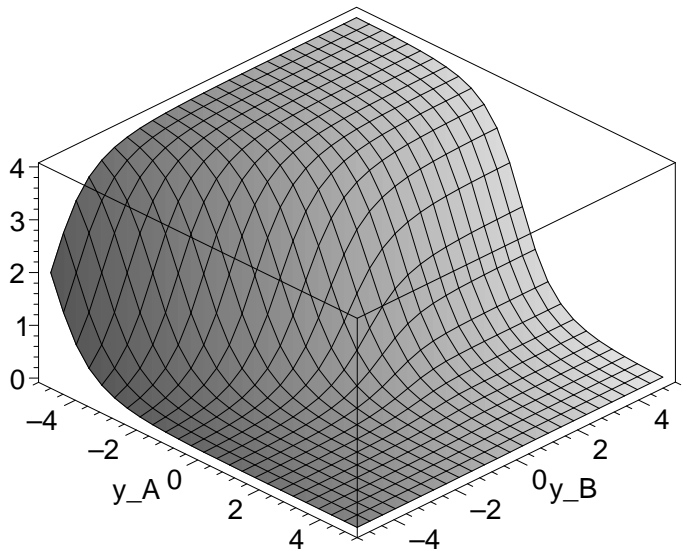}
\]
\end{figure}

Comparing Figure~3 with Figure~4, one can make a distinction
between four different regions in terms of the energy variables
$y_A$ and $y_B$.
The sector $(y_A<0, y_B>0)$ is populated mostly with particles
of kind~$A$, and the sector $(y_A>0, y_B<0)$ mostly with particles
of kind~$B$.
In the sector $(y_A<0, y_B<0)$, the population of particles
of kind~$A$ and of kind~$B$ is approximately the same.
Finally, the sector $(y_A>0, y_B>0)$ is essentially unpopulated.
The average number of accommodated particles is never bigger
that~4, as it should be, since $p=4$.

\section{Concluding remarks}
\setcounter{equation}{0}

In the present paper we have studied the thermal properties of
``free'' particles, which interact only via statistical interaction.
The latter stems from the restrictions imposed by the Pauli principle~:
the system under consideration cannot accommodate more that
$p$ particles if the order of statistics is~$p$. This property
holds independently of the number of orbitals; there can even be
infinitely many.

By definition $A$-statistics is closely related to certain
(more precisely, symmetric or Fock) representations of the Lie
algebra $sl(n+1)$, including $n=\infty$. Apart from that,
$A$-statistics belongs to the class of exclusion statistics as
defined in~\cite[Section~5]{Meljanac}.
Okubo~\cite{Okubo} has also reformulated this in the language
of Lie-triple systems. In~\cite{PV00} we have argued that under
certain natural assumptions $A$-statistics can be interpreted
as an exclusion statistics in the sense of Wu~\cite{Wu}.

Apart from the general case we have considered some specific
examples. In particular we have shown that for $n=1$ and any $p$
the FD distribution function ($n=1$, $p=1$) deforms into the BE distribution
function ($n=1$, $p=\infty$) with the growth of $p$,
see Figure~1. In this case $A$-statistics reduces to Gentile
statistics~\cite{Gentile} (see also~\cite{Chen}).
In the more general case of any number of orbitals~$n$
the above picture is modified. In the limit
$p \rightarrow \infty$ one obtains again the Bose distribution function
${\bar N}(p=\infty,n)=nx/(1-x)$. However at $p=1$ the distribution
function ${\bar N}(p=1,n)$ is a distribution function of hard-core
fermions, see~(\ref{3.46}), and not of fermions.

Another observation to mention is in the case with
equidistant energy levels. Without any input from quantum
groups it turns out that the GPF is a
$q$-deformation of the GPF of the most degenerate case.
More precisely, the equidistant GPF~(\ref{3.62}) is obtained
from the ``nondeformed" GPF~(\ref{3.31}) by a $q$-deformation of the
binomial coefficients. Another property natural to expect,
demonstrated here for $p=1$, is that at very low temperatures
the average number of particles of the system is the same as
the average number of particles on the lowest energy level,
which means that all allowed particles (in the general case $p$) 
``condensate" on the lowest level.

Despite of the fact that $A$-statistics does not belong to the class
of deformed Bose statistics, it yields a good approximation to
Bose statistics. Apart from that the Fock spaces do not contain
states with negative norm. Therefore, parallel to quons, $A$-statistics
with large values of $p$ is a good candidate for the description
of small violations of Bose statistics in quantum field theory.
Similarly as for quons~\cite{Greenberg2} however, we do not know
how to satisfy the locality condition in relativistic quantum field
theory. Therefore, one cannot expect to derive relations between
charge conjugation, unitarity and statistics as in~\cite{Cougo}.
It would be interesting to see whether such relations can be derived
in the frame of causal $A$-statistics~\cite{Palev80}.

Finally we point out that our considerations are incomplete in the
sense of traditional thermodynamics, because we have not introduced the
concept of volume and hence of pressure, etc. 
In our picture the volume
can be introduced in several ways. One natural way would be to relate
the order of statistics~$p$ to a unit volume $V$~: if $p$ is the maximal
number of particles to be accommodated in $V$, then it is
natural to assume that twice more particles could be accommodated in the
volume $2V$. This is one, but not the only plausible
possibility. We shall return to this
issue elsewhere.

\section*{Appendix~: Proof of equation~(\ref{3.69})}
\renewcommand{\theequation}{A.\arabic{equation}}
\setcounter{equation}{0}

First, we wish to find an expression for $h_k(x_1,\ldots,\hat x_i,
\ldots,x_n)$. The notation $\hat x_i$ means that $x_i$ has been
removed from the list of variables $(x_1,\ldots, x_n)$, so
$h_k(x_1,\ldots,\hat x_i, \ldots,x_n)$ stands for a symmetric function
in $n-1$ variables. 
Multiplying (\ref{3.8}) by $(1-x_it)$, it follows easily that
$$
h_k(x_1,\ldots,\hat x_i, \ldots,x_n) = 
h_k({\bf x}) - x_i h_{k-1}({\bf x}), 
$$
where $h_k({\bf x})\equiv h_k(x_1,x_2,\ldots,x_n)$.

Consider now the general expression for $\bar l_i$, 
as given in (\ref{3.20})~:
$$
\bar l_i = {1\over Z(p,n)}x_i \partial_{x_i} Z(p,n),
$$
or, using (\ref{3.5}),
\beas
x_i \partial_{x_i} Z(p,n)&=& \sum_{0\leq l_1+l_2+\cdots+l_n\leq p} l_i 
x_1^{l_1}x_2^{l_2}\cdots x_n^{l_n} \\
&=& \sum_{l_i=0}^p l_i x_i^{l_i} \sum_{0\leq l_1+\cdots+l_{i-1}+l_{i+1}
+\cdots+l_n\leq p} x_1^{l_1}\cdots x_{i-1}^{l_{i-1}}
x_{i+1}^{l_{i+1}}\cdots x_n^{l_n} \\
&=& \sum_{l_i=0}^p l_i x_i^{l_i}\sum_{k=0}^{p-l_i} 
\sum_{(l_1+\cdots+l_{i-1}+l_{i+1}+\cdots+l_n=p-k)} x_1^{l_1}\cdots x_{i-1}^{l_{i-1}}
x_{i+1}^{l_{i+1}}\cdots x_n^{l_n} \\
&=&\sum_{l_i=0}^p l_i x_i^{l_i}\sum_{k=0}^{p-l_i} 
h_k(x_1,\ldots,\hat x_i, \ldots,x_n)\\
&=&\sum_{l_i=0}^p l_i x_i^{l_i}\sum_{k=0}^{p-l_i} 
(h_k({\bf x})-x_i h_{k-1}({\bf x})).
\eeas
In this last expression, we can make the specialization $x_i=q^{i-1}x$.
{}From (\ref{3.61}) we know already how the functions $h_k({\bf x})$ specialize,
so there comes (replacing also the summation variable $l_i$ by $l$)
$$
\sum_{l=0}^p l (q^{i-1}x)^l \sum_{k=0}^{p-l} \left(
\left[n+k-1\atop k\right]x^k - q^{i-1}x \left[n+k-2\atop k-1\right]x^{k-1} \right).
$$
Replacing $q^{i-1}$ by a new variable $\alpha$, this can be rewritten as
$$
\sum_{l=0}^p \sum_{k=0}^{p-l} l \alpha^l \left(
\left[n+k-1\atop k\right] - \alpha \left[n+k-2\atop k-1\right]\right)
x^{k+l}.
$$
Collecting equal powers of $\alpha$, this reduces to
\beq
\sum_{l=1}^p \alpha^l \sum_{k=0}^{p-l} \left[n+k-1\atop k\right] x^{k+l}.
\label{a1}
\eeq
Putting back $\alpha=q^{i-1}$ gives the relation (3.69), which
we wanted to prove.

Observe that one summation can be performed in (\ref{a1})~:
\beas
\sum_{l=1}^p \alpha^l \sum_{k=0}^{p-l} \left[n+k-1\atop k\right] x^{k+l}&=&
\sum_{k=0}^{p-1} \left[n+k-1\atop k\right] x^{k} \sum_{l=1}^{p-k} 
(\alpha x)^l \\
&=&\sum_{k=0}^{p-1} \left[n+k-1\atop k\right] x^{k} 
\bigl({\alpha x -(\alpha x)^{p-k+1} \over 1-\alpha x}\bigr).
\eeas
Replacing again $\alpha$ by $q^{i-1}$ yields an alternative
expression for (\ref{3.69}).

\section*{Acknowledgements}
The authors would like to thank the referee for pointing out
some relevant references.
T.D.\ Palev was supported by NATO (Collaborative
Linkage Grant) during his visit to Ghent.
He also wishes to acknowledge Ghent University for a
visitors grant.
A.\ Jellal and T.D.\ Palev are grateful to Prof.\ Randjbar-Daemi
for the kind hospitality at the High Energy Section
of ICTP, Trieste, where part of this work was initiated.

\newpage
\listoffigures


\newpage
\begin{thebibliography}{99}

\bibitem{Gentile}
G.\ Gentile, 
{\em Nuov.\ Cim.} {\bf 17}, 493 (1940).

\bibitem{Wigner}
E.P.\ Wigner,
{\em Phys.\ Rev.} {\bf 77}, 711 (1950).

\bibitem{Green}
H.S.\ Green, 
{\em Phys.\ Rev.} {\bf 90}, 270 (1953).

\bibitem{Greenberg}
O.W.\ Greenberg,
{\em Phys. Rev.} {\bf D 43}, 4111 (1991).

\bibitem{Wilczek}
F.\ Wilczek,
{\em Phys.\ Rev.\ Lett.} {\bf 48}, 1144 (1982).

\bibitem{Haldane}
F.D.M.\ Haldane,
{\em Phys.\ Rev.\ Lett.} {\bf 67}, 937 (1991).

\bibitem{Wu}
Y.-S.\ Wu, 
{\em Phys.\ Rev.\ Lett.} {\bf 73}, 922 (1994).

\bibitem{Berkovich}
A.\ Berkovich and B.M.\ McCoy,
``The universal chiral partition function for exclusion statistics''
(Preprint hep-th/9808013).

\bibitem{Pusz}
W.\ Pusz and S.L.\ Woronowicz,
{\em Rep.\ Math.\ Phys.} {\bf 27}, 231 (1989);
{\bf 27}, 349 (1989).

\bibitem{Macfarlane}
A.J.\ Macfarlane,
{\em J.\ Phys.\ A~: Math.\ Gen.} {\bf 22}, 4581 (1989);\\
L.C.\ Biedenharn,
{\em J.\ Phys.\ A~: Math.\ Gen.} {\bf 22}, L873 (1989);\\
C.P.\ Sun and H.C.\ Fu,
{\em J.\ Phys.\ A~: Math.\ Gen.} {\bf 22}, L983 (1989).

\bibitem{Bonatsos}
D.\ Bonatsos, C.\ Daskaloyannis and P.\ Kolokotronis,
{\em Mod.\ Phys.\ Lett.} {\bf A 10}, 2197 (1995),
and preprint hep-th/9512083.

\bibitem{Mishra}
A.K.\ Mishra and G.\ Rajasekaran,
{\em Pramana J.\ Phys.} {\bf 45}, 91 (1995).

\bibitem{OK}
Y.\ Ohnuki and S.\ Kamefuchi,
{\em Quantum Field Theory and Parastatistics}, 
Springer, Berlin, 1982;\\
T.D.\ Palev,
{\em J.\ Math.\ Phys.} {\bf 23}, 1778 (1982).

\bibitem{Palev76}
T.D.\ Palev,
``Lie algebraical aspects of the quantum statistics''
(Habilitation thesis, Inst.\ Nuclear Research \& Nucl.\
Energy, Sofia, 1976; in Bulgarian).

\bibitem{Palev77}
T.D.\ Palev, 
``Lie algebraical aspects of quantum statistics. 
Unitary quantization (A-quantization)'' (Preprint JINR E17-10550, 1977; 
preprint hep-th/9705032).

\bibitem{PV00}
T.D.\ Palev and J.\ Van der Jeugt,
``Jacobson generators, Fock representations
and statistics of $sl(n+1)$'' (Preprint hep-th/0010107).

\bibitem{KT}
S.\ Kamefuchi and Y.\ Takahashi,
{\em Nucl.\ Phys.} {\bf 36}, 177 (1962).

\bibitem{RS}
C.\ Ryan and E.C.G.\ Sudarshan, 
{\em Nucl.\ Phys.} {\bf 47}, 207 (1963).

\bibitem{Karabali}
D.\ Karabali and V.P.\ Nair,
{\em Nucl.\ Phys.\ B} {\bf 438}, 551 (1995).

\bibitem{Liguori}
A.\ Liguori and M.\ Mintchev,
{\em Lett.\ Math.\ Phys.} {\bf 33}, 283 (1995).

\bibitem{Jacobson}
N.\ Jacobson,
{\em Amer.\ J.\ Math.} {\bf 71}, 149 (1949).

\bibitem{Poly}
A.P.\ Polychronakos,
{\em Nucl.\ Phys.\ B} {\bf 324}, 597 (1989).

\bibitem{Macdonald}
I.G.\ Macdonald, 
{\em Symmetric Functions and Hall polynomials},
Clarendon Press, Oxford, 1995.

\bibitem{Greenberg2}
O.W.\ Greenberg,
``Theories of violation of statistics''
(Preprint hep-th/0007054).

\bibitem{Chaichian}
M.\ Chaichian, R.G.\ Felipe and C.\ Montonen,
{\em J.\ Phys.\ A} {\bf 26}, 4017 (1993).

\bibitem{Slater}
L.J.\ Slater, 
{\em Generalized Hypergeometric Functions},
Cambridge University Press, Cambridge, 1966.

\bibitem{GR}
G.\ Gasper and M.\ Rahman,
{\em Basic Hypergeometric Series},
Cambridge Univ.\ Press, Cambridge, 1990.

\bibitem{Meljanac}
S.\ Meljanac, M.\ Milekovic and M.\ Stojic,
{\em J.\ Phys.\ A} {\bf 32}, 1115 (1999).

\bibitem{Okubo}
S.\ Okubo,
{\em J. Math.\ Phys.} {\bf 35}, 2785 (1994 ).

\bibitem{Chen}
W.\ Chen, Y.\ Jack Ng and H.\ Van Dam,
{\em Mod.\ Phys.\ Lett.\ A} {\bf 11}, 795 (1996).

\bibitem{Cougo}
M.V.\ Cougo-Pinto,
{\em Phys.\ Rev.\ D} {\bf 46}, 858 (1992).

\bibitem{Palev80}
T.D.\ Palev,
{\em Rep.\ Math.\ Phys.} {\bf 18}, 117 (1980).

\end{thebibliography}
\end{document}